\documentclass[10pt,letterpaper]{article}
\usepackage{opex3}
\usepackage{epsfig}
\usepackage{amsmath}
\usepackage{amssymb}
\usepackage{mathrsfs}

\begin{document}

\title{The radiated fields of the fundamental mode of photonic crystal fibers}

\author{Ali Dabirian,$^1$ Mahmood Akbari,$^1$ and Niels Asger Mortensen$^2$}

\address{$^1$Department of Electrical Engineering,~Sharif
University of Technology,\\ Tehran, 14115-337, Iran\\
$^2$MIC -- Department of Micro and Nanotechnology, Technical
University of Denmark, DK-2800 Kongens Lyngby, Denmark}

\email{ali\_dabirian@yahoo.com}

\date{January 26, 2005}

\begin{abstract}The six-fold rotational symmetry of photonic
crystal fibers has important manifestations in the radiated fields
in terms of {\it i)} a focusing phenomena at a finite distance
from the end-facet and {\it ii)} the formation of low-intensity
satellite peaks in the asymptotic far field. For our study, we
employ a surface equivalence principle which allows us to
rigorously calculate radiated fields starting from fully-vectorial
simulations of the near field. Our simulations show that the
focusing is maximal at a characteristic distance from the
end-facet. For large-mode area fibers the typical distance is of
the order $10\times\Lambda$ with $\Lambda$ being the pitch of the
triangular air-hole lattice of the photonic crystal fiber.
\end{abstract}



The photonic crystal fiber~\cite{knight1996,birks1997} (PCF)
offers a unique opportunity for realizing broadband large-mode
area (LMA) single-mode (SM) fibers~\cite{nielsen2004a}. In
high-power laser applications the LMA property is desirable to
keep the power density below the damage threshold of silica as
well as to avoid non-linear phenomena and the SM property is
essential for the beam quality and the ability to generate close
to diffraction-limited beams. Recent studies of LMA-SM-PCF lasers
have suggested that the PCF technology may indeed offer good beam
quality~\cite{limpert2003} as quantified by an M-squared
value~\cite{johnston1990} close to unity ($M^2=1$ for a Gaussian
beam).

For the fundamental mode in PCFs the near and far fields have
intensity distributions which are overall close to Gaussian, but
with a six-fold rotational symmetry reflecting the symmetry of the
air-hole lattice. For LMA-PCFs there are no significant deviations
from a Gaussian distribution for intensities above the $1/e^2$
level~\cite{mortensen2002a} and for many practical applications
the mode may be considered Gaussian. However, as studied recently
the derivation from a Gaussian distribution has some important
consequences~\cite{mortensen2002b}. In the near-field the
intensity is suppressed at the locations of the air-holes and in
the asymptotic far-field these suppressions manifest themselves as
low-intensity satellite peaks. Furthermore, in the near field to
far-field transition the intensity becomes strongly focused at a
finite distance from the end-facet as can be understood by a
simple Gaussian decomposition of the near field.

While the simple decomposition approach has proved very useful for
the qualitative understanding of the near to far-field transition
the laser applications call for a more quantitative study of the
radiated field. Calculating the diffraction pattern is generally a
complicated task, which involves the full solution to the elastic
scattering problem at the end-facet. Here, we rigorously calculate
radiated fields from the end-facet utilizing a surface equivalence
principle~\cite{balanis1989}.

The calculations of the modal reflectivity from the fiber
end-facet is performed with a three-dimensional frequency domain
modal method relying on Fourier expansion
technique~\cite{silberstein2001,cao2002}. In brief, the Fourier
expansion method relies on an analytical integration of Maxwell's
equations along one direction (the fiber axis) and on a supercell
approach in the other two transversal directions. We emphasize
that for the radiated field at the surface $S_1$ we have treated
the elastic scattering at the end-facet fully, i.e. the incident
fundamental mode is partly reflected (no mode mixing for symmetry
reasons) and partly radiated/transmitted. Through the surface
equivalence principle, any wave-front can be considered as a
source of secondary waves which add to produce distant wave-fronts
according to Huygens principle.

\begin{figure}[t!]
  \centering
  \includegraphics*[angle=0, width=0.8\columnwidth]{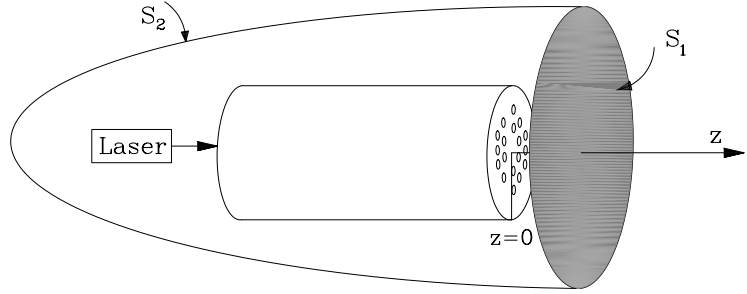}
\caption{Schematics of the fiber geometry and the imaginary closed
surface $S=S_1+S_2$, where $S_1$ is a circle parallel and close to
the end-facet and $S_2$ is a semi-sphere concentric with $S_1$.
}\label{fig1}
\end{figure}

In order to apply the surface equivalence principle, we surround
the end-facet radiating fiber with an imaginary closed surface
$S=S_1+S_2$, where $S_1$ is a circle parallel and close to the
end-facet and $S_2$ is a semi-sphere concentric with $S_1$, see
Fig.~\ref{fig1}. Both of these surfaces are of infinite extent. By
considering proper surface currents on $S$ which are equivalent in
the sense that they yield the same field outside $S$, we can
formally erase all the existing elements inside $S$ and fill it
with a perfect electric conducting (PEC) body. The equivalent
surface electrical and magnetic currents on $S$ are ${\mathbf
J_e}({\mathbf r})= {\mathbf {\hat n}}\times {\mathbf H }$ and
${\mathbf M_e}({\mathbf r})=- {\mathbf {\hat n}}\times {\mathbf E
}$, respectively, where ${\mathbf {\hat n}}$ is an outward normal
to the surface $S$~\cite{balanis1989}. Of course, these currents
radiate in the presence of the PEC and not in unbounded
free-space. Because of the infinite radius of $S_2$,
electromagnetic fields on this surface take zero value. On $S_1$
electromagnetic fields are determined from near-fields. Since the
$S_1$ surface is an infinitely extended flat PEC body we can
utilize image theory to replace the conductor by image currents
${\mathbf {-J_e}}$ and ${\mathbf M_e}$. Ultimately, computation of
radiated fields of an open-ended PCF is simplified to computing
radiated fields of the fictitious current source ${\mathbf
M}({\mathbf r})=2{\mathbf M_e}({\mathbf r})=-2  {\mathbf {\hat
n}}\times {\mathbf E}$ in free-space. Calculating radiated fields
of the current source can be achieved systematically through
electric ${\mathbf \pi}_e({\mathbf r})$ and magnetic ${\mathbf
\pi}_m({\mathbf r})$ Hertz vector potentials defined by:
\begin{subequations}
\begin{align}
 {\mathbf \pi}_e({\mathbf r}) &= \frac{-j}{4\pi\omega\epsilon_0} \int_{S}
 {\mathbf J}({\mathbf r}')
 \frac{e^{-jk\mid {\mathbf r}-{\mathbf r}' \mid}}{\mid {\mathbf r}-{\mathbf r}'\mid}
 \:{dS'},
 \label{g1}\\
{\mathbf \pi}_m({\mathbf r}) &= \frac{-j}{4\pi\omega\mu_0}
\int_{S} {\mathbf M}({\mathbf r}') \frac{e^{-jk\mid {\mathbf
r}-{\mathbf r}'\mid}}{\mid {\mathbf r}-{\mathbf r}'\mid}\: {dS'}.
\label{g2}
\end{align}
\end{subequations}
Employing these potentials, electric and magnetic fields are
obtained by:
\begin{subequations}
\begin{align}
{\mathbf E}({\mathbf r})&=(k^2+\nabla^2){\mathbf \pi}_e({\mathbf
r}) -j\omega\mu_0\nabla\times {\mathbf \pi}_m({\mathbf r}),
 \label{g3}\\
{\mathbf H}({\mathbf r})&=j\omega\epsilon_0\nabla\times{\mathbf
\pi}_e({\mathbf r})+
 (k^2+\nabla^2) {\mathbf \pi}_m({\mathbf r}),
 \label{g4}
\end{align}
\end{subequations}
where $\omega=ck$, $k=2\pi/\lambda$, and $\lambda$ is the
free-space wavelength. From Eqs.~(\ref{g3}) and (\ref{g4}) we
expect an improvement in computational time by a factor of roughly
two compared to a direct use of the surface equivalence
principle~\cite{vuckovic2002} in which both fictitious electric
and magnetic surface currents exist.

\begin{figure}[t!]
  \centering
  \includegraphics*[angle=0, width=0.8\columnwidth]{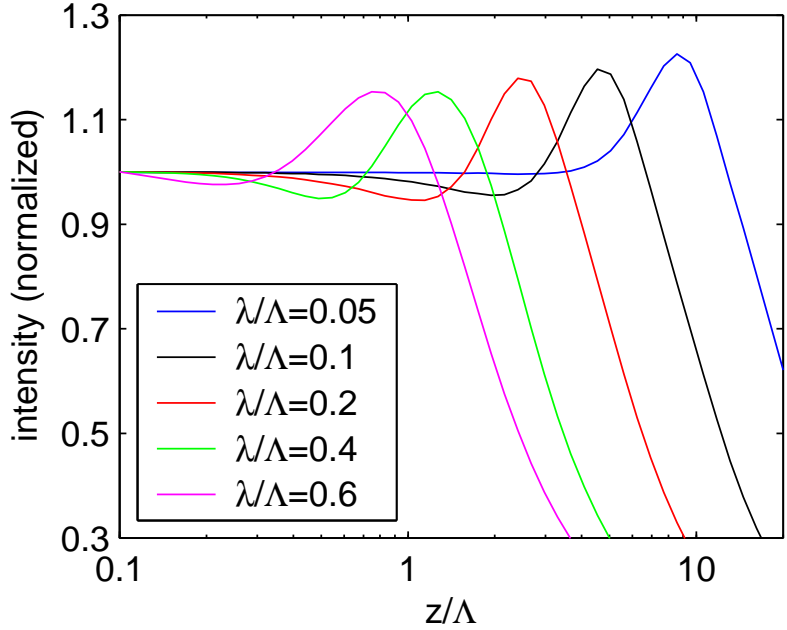}
\caption{ Variation of the electric field intensity with distance
from the end-facet ($z=0$) at the center of fiber.}\label{fig2}
\end{figure}

We employ the described approach in a fully vectorial study of the
evolution of the fundamental mode of a PCF from near-field to
far-field. As an example we consider the fundamental mode in a
pure-silica PCF with a normalized air-hole diameter
$d/\Lambda=0.45$ and in our simulations we utilize the
scale-invariance of Maxwell's equations for a
frequency-independent dielectric function. Recent experimental
studies showed quite complicated interference phenomena in the
near to far-field transition~\cite{mortensen2002b} and in
particular a focusing behavior was observed at a finite distance
from the end-facet. As illustrated in Fig.~\ref{fig2} this
focusing phenomena is born out by our numerical simulation where
the peak intensity of the radiation increases up to short
distances from the fiber end-facet. The electric field intensity
is maximal at a distance $z_0$ which is typically in the range
$\Lambda \lesssim z_0 \lesssim 10\times\Lambda$. At this
particular distance the intensity pattern is rotated by an angle
of $\pi/6$ compared to the near-field, in full agreement with the
experimental observations. As the wavelength is increased relative
to the pitch the diffraction by the six inner-most air holes
increases and $z_0$ shifts closer to the end facet of the PCF. We
emphasize that this interference phenomenon has no counterparts
for e.g. step-index fibers with cylindrical symmetry; the focusing
relies fully on the six-fold rotational symmetry of the air-hole
lattice. It should also be emphasized that the phase-front is
non-trivial (non-constant) in the transverse plane at $z_0$ and in
that sense the focusing differs from that observed with a
classical lens.

\begin{figure}[t!]
  \centering
  \includegraphics*[angle=0, width=\columnwidth]{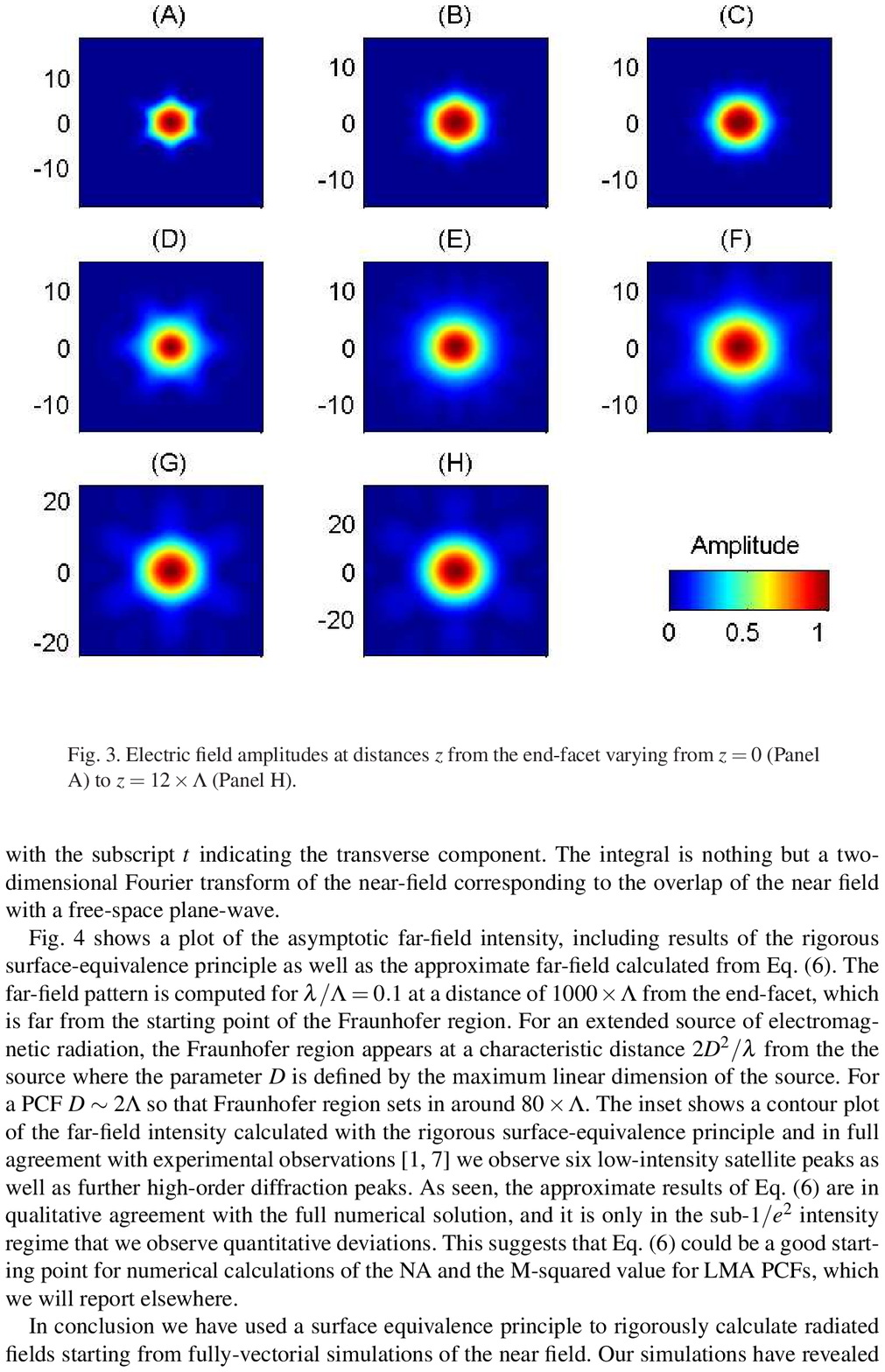}
\caption{Electric field amplitudes at distances $z$ from the
end-facet varying from $z=0$ (Panel A) to $z=12\times\Lambda$
(Panel H).}\label{fig3}
\end{figure}

In Fig.~\ref{fig3} we show the electric field amplitude at various
distances from the end-facet up to the asymptotic far-field
region. The nearly hexagonal shape of the fundamental mode of the
fiber (Panel A) is transformed into a nearly circular shape (Panel
B) which is followed by a hexagonal shape at $z\simeq z_0$ (Panel
C) where the orientation is rotated by an angle of $\pi/6$ with
respect to the fiber mode at $z=0$. This hexagonal shape expands
as it propagates (Panels D and E) and then becomes close to
circular again above the $1/e$ amplitude level (Panel F). As the
distance from the end-facet becomes larger, six satellites start
emerging from the mode shape below the $1/e$ amplitude level
(Panel G). Finally, for larger distances satellites appear clearly
(Panel H).

As an alternative to the above rigorous approach, it is in fiber
optics common to approximate the far field by a Fourier
transformation of the near field~\cite{ghatak1998}. Here we shall
derive this expression starting from the above formalism. In the
far-field limit we approximate the $\big|{\mathbf r}-{\mathbf
r}'\big|$ in the denominators of Eqs.~(\ref{g1}) and (\ref{g2}) by
$r$ and in the exponential we correspondingly approximate it by
$r-{\mathbf {\hat r}}\cdot{\mathbf r}'$, where ${\mathbf {\hat
r}}={\mathbf r}/r$ is the unit radial vector of the spherical
coordinate. The magnetic Hertz vector potential, Eq.~(\ref{g2}),
then simplifies to
\begin{align}
{\mathbf \pi}_m({\mathbf r}) \simeq \frac{-j}{4\pi\omega\mu_0}
\frac{\exp(-jkr)}{r}\:{\mathbf N}(\theta,\phi), \label{g5}
 \end{align}
where
\begin{align}
{\mathbf N}({\theta,\phi}) = \int_{S}
 {\mathbf M}({\mathbf r}')
 {e^{jk{\mathbf {\hat r}}\cdot{\mathbf r}'}}  \:{dS'},\label{g6}
 \end{align}
with $\theta$ and $\phi$ being the polar and azimuthal angles,
respectively, and ${\mathbf r}=r{\mathbf {\hat r}}+\phi{\mathbf
{\hat \phi}}+\theta{\mathbf {\hat \theta}}$. The electric field
now becomes
\begin{equation}
{\mathbf E}({\mathbf r})\simeq
\frac{-1}{4\pi}\frac{e^{-jkr}}{r}(-jk)({\mathbf {\hat
r}}\times{\mathbf N}).\label{g7}
\end{equation}
We will consider small angles $\theta$ of divergence so that
${\mathbf {\hat r}}\times{\mathbf M} = -M_y  {\mathbf {\hat x}}
+M_x{\mathbf {\hat y}} +{\cal O}(\theta)$ by which Eq.~(\ref{g7})
simplifies to
\begin{equation}\label{g8}
{\mathbf E}({\mathbf r})\simeq
\frac{jk}{2\pi}~\frac{e^{-jkr}}{r}\int_{S} {\mathbf
E}_t(x',y',z'=0)  \times \exp\big[{jk(x' sin\theta cos\phi+ y'
sin\theta sin \phi})\big] \:{dS'},
\end{equation}
with the subscript $t$ indicating the transverse component. The
integral is nothing but a two-dimensional Fourier transform of the
near-field corresponding to the overlap of the near field with a
free-space plane-wave.

\begin{figure}[t!]
  \centering
  \includegraphics*[angle=0, width=0.8\columnwidth]{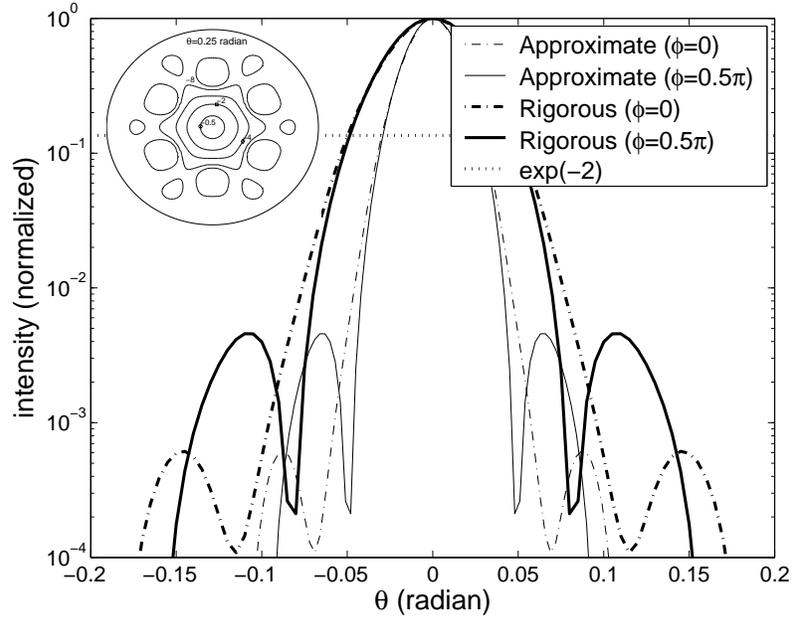}
\caption{Asymptotic far-field intensities calculated by the
rigorous approach as well as the approximate expression,
Eq.~(\ref{g8}), along the high-symmetry directions $\phi=0$ and
$\phi=\pi/2$. The inset shows the corresponding full contour plot
obtained with the rigorous approach.}\label{fig4}
\end{figure}

Fig.~\ref{fig4} shows a plot of the asymptotic far-field
intensity, including results of the rigorous surface-equivalence
principle as well as the approximate far-field calculated from
Eq.~(\ref{g8}). The far-field pattern is computed for
$\lambda/\Lambda=0.1$ at a distance of $1000\times \Lambda$ from
the end-facet, which is far from the starting point of the
Fraunhofer region. For an extended source of electromagnetic
radiation, the Fraunhofer region appears at a characteristic
distance $2D^2/\lambda$ from the source where the parameter $D$ is
defined by the maximum linear dimension of the source. For a PCF
$D\sim 2\Lambda$, so Fraunhofer region sets in around
$80\times\Lambda$. The inset shows a contour plot of the far-field
intensity calculated with the rigorous surface-equivalence
principle. We observe six low-intensity satellite peaks as well as
further high-order diffraction peaks, in full agreement with
experimental observations~\cite{knight1996,mortensen2002b}. As
seen, the approximate results of Eq.~(\ref{g8}) are in qualitative
agreement with the full numerical solution, and it is only in the
sub-$1/e^2$ intensity regime that we observe quantitative
deviations. This suggests that Eq.~(\ref{g8}) could be a good
starting point for numerical calculations of the NA and the
M-squared value for LMA PCFs, which we will report elsewhere.

In conclusion we have used a surface equivalence principle to
rigorously calculate radiated fields starting from fully-vectorial
simulations of the near field. Our simulations have revealed a
focusing behavior which is maximal at a characteristic distance,
of the order $10\times\Lambda$ from the end-facet. In the
far-field limit we have shown how qualitative and to some degree
also quantitative insight may be gained from simple
two-dimensional Fourier transforms of the near-field.

\vspace{5mm} We thank J.~R. Folkenberg and P.~M.~W. Skovgaard
(both with Crystal Fibre A/S) for stimulating discussions.
N.~A.~M. is supported by The Danish Technical Research Council
(Grant No.~26-03-0073).

\end{document}